\documentstyle[aps,preprint,epsf]{revtex}  
\tighten
\begin{document}
\title{Critical dynamics and multifractal exponents at the Anderson 
transition in 3d disordered systems} 
\author{$^1$T. Brandes, $^2$B. Huckestein and $^3$L. Schweitzer}
\address{$^1$Department of Physics, Gakushuin University, 1-5-1 Mejiro, 
Toshima-ku, Tokyo 171, Japan\\
$^2$Institut f\"ur Theoretische Physik, Universit\"at zu K\"oln, 
Z\"ulpicher Str. 77, 50937 K\"oln, Germany\\
$^3$Physikalisch-Technische Bundesanstalt, Bundesallee 100, 
38116 Braunschweig, Germany}
\date{\today} 
\maketitle
\begin{abstract}
We investigate the dynamics of electrons in the vicinity of the Anderson 
transition in $d=3$ dimensions. Using the exact eigenstates from a numerical 
diagonalization, a number of quantities related to the critical behavior of 
the diffusion function are obtained. The relation $\eta = d-D_{2}$ between 
the correlation dimension $D_{2}$ of the multifractal eigenstates  
and the exponent $\eta$ which enters into correlation functions is verified. 
Numerically, we have $\eta\approx 1.3$. Implications of critical
dynamics for experiments are predicted. We investigate the long-time
behavior of the motion of a wave packet. Furthermore, electron-electron and 
electron-phonon scattering rates are calculated.
For the latter, we predict a change  of the temperature dependence for low
$T$  due to $\eta$. The electron-electron scattering rate is found to be
linear in $T$ and depends on the dimensionless conductance at the critical
point.
\end{abstract}

\section{Introduction} 

The dynamics of non-interacting electrons in the 
vicinity or directly at the critical point of a metal-insulator transition is 
still an unsolved problem. In the last years, a number of works 
\cite{Weg80,SE84,PJ91,HS92,SG91,Jan94,HS94,Sch95}
revealed strong amplitude fluctuations of the eigenstates near the 
critical energy $E_{c}$. The wave functions turned out to be multifractal 
objects, described by a set of generalized fractal dimensions $D_{q}$. 
On the other 
hand, investigation of correlation functions \cite{CD88} in combination with 
scaling arguments showed that on certain length and time scales the dynamics
near $E_{c}$ is governed by anomalous diffusion, 
described by an exponent $\eta$. One possible definition \cite{etadefinition}
for $\eta$ is the
algebraic decay \cite{Weg76} of the static two-particle 
correlation function $S(r,E,\omega \to 0)$ near the critical energy 
$E\to E_{c}$
\begin{equation} 
\label{Somega0}
S(r,E,\omega\rightarrow 0 )\sim \left ( \frac{r}{\xi(E)} \right )^{-\eta} ,
\quad r \ll \xi(E)
\end{equation} 
where $\xi(E)$ is the localization length. In real space, $S$ is defined as
\begin{equation}
\label{Sdefinition}
S(r,E,\omega):= \left\langle
\sum_{\alpha ,\beta }\delta (E^- -E_{\alpha }) \delta (E^+ -E_{\beta})
\Psi_{E^-}(0) \Psi_{E^-}^*(r)\Psi_{E^+}(r)\Psi_{E^+}^*(0)\right\rangle,
\end{equation}
where $E^{\pm}:=E\pm\hbar\omega /2$, $\alpha ,\beta $ label the 
eigenstates, and the brackets $\langle\phantom{a}\rangle$ denote an impurity 
average. 
The correlation function $S$ is connected to the diffusion function 
$D(q,\omega )$ via 
\begin{equation}
\label{sd}
S(q, E, \omega )=\frac{\rho(E)}{\pi\hbar}\frac{D(q,\omega )q^2}
{\omega ^2  +\left ( D(q,\omega )q^2\right )^2 },
\end{equation}
where in turn $D(q,\omega )$ is the generalization of the diffusion
constant $D$ in the metallic case and $\rho(E)$ is the density of states.
Indeed,  Eq.~(\ref{sd}) is a general expression for the two-particle 
correlation function compatible with particle conservation \cite{Forster}.
The function $D(q,\omega )$ appears in the diffusion propagator
\begin{equation}
\label{diffusionpropagator}
P(q,\omega):=\frac{1}{-i\omega +D(q,\omega )q^2},
\end{equation}
which is the Fourier transform of the probability distribution $P({\bf r},t)$ 
describing the motion of an electron wave packet in a disordered system for 
$t>0$, initially located at ${\bf r}=0$ at time $t=0$. 

In this work, we determine the correlation dimension $D_2$ of the multifractal
eigenstates and the exponent $\eta$ for the Anderson 
transition in 3 dimensions. The relation 
\begin{equation} 
\label{relation} 
D_{2}=d-\eta, 
\end{equation} 
which holds in $d$ dimensions, is derived from the respective definitions of 
the exponents and verified numerically. We independently determine $D_{2}$ 
via the box counting method applied to the spatial amplitude fluctuations of 
the critical eigenstates on the one side, and $\eta$ from a correlation 
function in energy space using a scaling form of $D(q,\omega )$ on the other 
side. Furthermore, the relation $\delta =\widetilde{D_{2}}=D_{2}/d$ is 
established and checked numerically, where $\delta $ governs the probability 
of return of a wave packet, $p(t)\sim t^{-\delta }$, and $\widetilde{D_{2}}$ 
is the correlation dimension of the spectral measure at the critical point. 

Two applications which might be relevant for experiments are presented. 
First, we discuss the long time behavior of a wave packet of eigenstates with 
energies near $E_{c}$ and show that in three dimension this behavior is 
drastically different from its two-dimensional counterpart \cite{HS94,Sch95}. 
Then, 
it is shown how rates for inelastic electron-electron and electron-phonon 
scattering are changed due to the critical dynamics as compared with the 
usual metallic case. We conclude with an overview of the results for various 
exponents in the two dimensional quantum Hall, the two dimensional 
symplectic, and the three dimensional orthogonal case. 

\section{Scaling form of \mbox{$D(q,\omega )$}}

The one-parameter scaling hypothesis \cite{AALR79}, which underlies a great 
part of the theoretical analysis of the metal-insulator transition
\cite{KM94}, allows to 
write the diffusion function $D(q,\omega)$ in $d$ dimensions near the 
critical point in the form \cite{AL86} 
\begin{equation} D(q,\omega )=\frac{\xi^{2-
d}}{\rho}F\left(\xi/L_\omega,\xi q\right) 
\end{equation} 
where $\xi$ is the localization length, $\rho=\rho (E)$ the density of
states (DOS) near the critical energy $E=E_{c}$, $F$ is a scaling
function, and
\begin{equation}
\label{lw}
L_{\omega } := (\rho \hbar \omega )^{-1/d}=L(\hbar\omega /\Delta )^{-1/d},
\end{equation}
is a third length scale, besides $q^{-1}$ and $\xi$, relevant at the
critical point ($\Delta =(L^{d}\rho )^{-1}$ is the mean level
spacing).

In addition to $L_{\omega}$, one can define a second frequency
dependent length
\begin{equation}
  \label{lwtilde}
  \widetilde{L_{\omega}}= \left( \frac{D(\omega)}{\omega}\right)^{1/2},
\end{equation}
the distance the particle diffuses in the time $1/\omega$. Here
$D(\omega)=\lim_{q\to0}D(q,\omega)$ is explicitly
scale-dependent. Apparently, $L_{\omega}$ and $\widetilde{L_{\omega}}$
scale with different powers of the frequency. However, at the mobility
edge this not the case. From the definition eq.~(\ref{lwtilde}), the
dimensionless conductance for a hypercube of size
$\widetilde{L_{\omega}}$ is given by
\begin{equation}
  \label{ohm}
  g^*=G\frac{\hbar}{e^2}=\frac{\hbar}{e^2}\sigma(\omega)\widetilde{L_{\omega}}^{d-2}. 
\end{equation}
Here, the frequency-dependent conductivity $\sigma(\omega)$ is related
to the diffusion coefficient by the Einstein relation
\begin{equation}
  \label{einstein}
  \sigma(\omega)=e^2 \rho D(\omega).
\end{equation}
Eliminating $D(\omega)$ from these equations leads to
\begin{equation}
  \label{lwtildesol}
  \widetilde{L_{\omega}} = \left(\frac{g^*}{\hbar \rho
      \omega}\right)^{1/d} = {g^*}^{1/d}L_{\omega}.
\end{equation}
Thus, at the mobility edge both frequency-dependent length scales
differ only by a factor of ${g^*}^{1/d}$.

The general 
relation between the conductivity $\sigma$ and the conductance $G$ for a 
sample in the form of a hypercube with sides of length $L$, $G(L)=\sigma(L) 
L^{d-2}$, and the vanishing of the $\beta $-function $\partial G/\partial L 
=0$  at $E_{c}$ \cite{AALR79} requires the conductivity and therefore the 
diffusion function $D\sim \sigma$ to scale like 
\begin{equation} 
D\sim L^{2-d} 
\end{equation} 
at the critical point. 

\begin{figure}[htb]
\unitlength1.0cm
\begin{picture}(12,10)
\put(-2,-19){\epsfbox{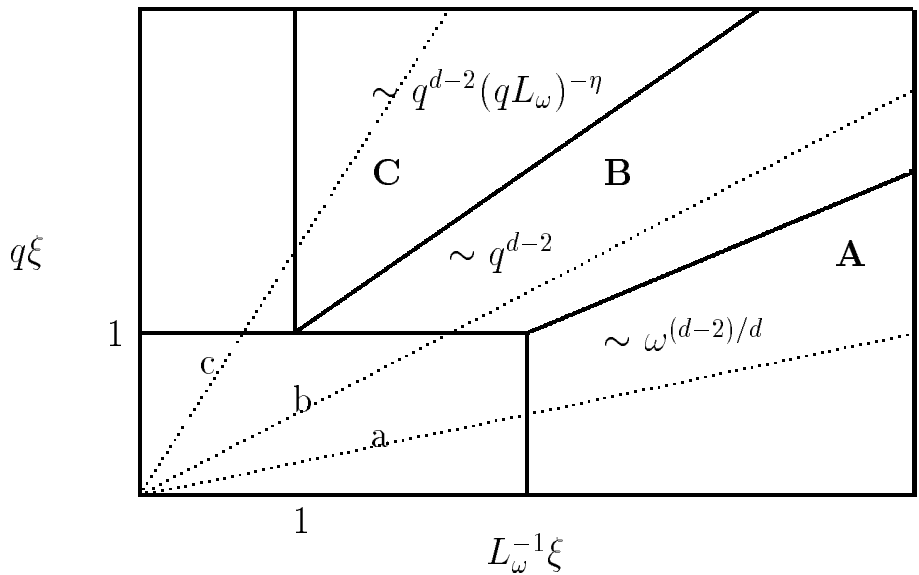}}
\end{picture}  
\caption[]{\label{eta3}
Five different regimes (separated by solid lines) for the 
diffusion function $D(q,\omega)$ after Chalker \cite{Cha90}. Relevant for 
our numerical analysis are the three regimes A,B,C ($\xi \to \infty$) with  
the behavior of $D(q,\omega ) \sim {\omega}^{(d-2)/d}$, $\sim q^{d-2}$, 
and in particular $q^{d-2}(qL_{\omega })^{-\eta}$ 
with the exponent $\eta=D_{2}-d$ related to the 
multifractality of the eigenstates. According to the value of the scaling 
variable $x=(qL_{\omega })^{d}$,  the function $f(x)$, Eq.~(\ref{fx}), follows 
a different power-law in $x$. The dotted lines indicate this crossover between 
A, B and C. Since $q\xi= x^{1/d}(L_{\omega }^{-1}\xi)$, (a) 
corresponds to small, (b) to intermediate and (c) to large $x$. The latter 
case is relevant for the numerical determination of $\eta$.} 
\end{figure}

A direct consequence of this scaling is a certain behavior of
$D(q,\omega )$ in different regions of the $(q,\omega)$-plane, as
discussed by Chalker \cite{Cha90}. In the limit of small $q \to 0$,
namely $qL_{\omega }\ll 1$, $L_{\omega }$ sets the shortest length
scale (region (A) in Fig.~\ref{eta3}), and
\begin{eqnarray}
D(q\to 0,\omega )\sim L_\omega^{2-d}\sim \omega^{(d-2)/d},\quad \omega
\rightarrow 0. 
\end{eqnarray}
This behavior has been predicted by 
Wegner in 1976 \cite{Weg76}. Recently, $D(\omega )\sim \omega ^{1/3}$ has 
been confirmed numerically for  $d=3$ \cite{LS94}. 

For larger $q$ (region (B)), the length scale
determining the scaling of $D(q,\omega)$ is the inverse wave vector itself,
if the localization length is not shorter than $1/q$. This leads to
\begin{equation}
\label{D(q,0)}
D(q,\omega)\sim  q^{d-2}.
\end{equation}
It has turned out in the last years that there are corrections to the above 
scaling form due to strong fluctuations of the wave function amplitudes near 
the critical point. These fluctuations are observed numerically on small 
length scales corresponding to large $q$-vectors. In the $(q,\omega )$-plane, 
they give rise to another region (C) where $D(q,\omega )$,
Eq.~(\ref{D(q,0)}), 
acquires an additional factor $q^{-\eta}$, the scale of which is given by 
either the localization length $\xi $ or  the length $L_{\omega }$, namely 
\begin{equation} 
\label{D} 
D(q,\omega )\sim 
q^{d-2}\left[q\cdot{\rm min}(\xi,L_{\omega }) \right]^{-\eta}.
\end{equation} 
For our numerical analysis, we use the following scaling form of the 
diffusion function: 
\begin{equation}
D(q,\omega )=\frac{q^{d-2}}{\hbar\rho }f(x),\qquad x:=\left[
q\cdot{\rm min}(\xi,L_{\omega }) \right]^d,
\end{equation}
which  satisfies the scaling relation ($\xi(E)>L_{\omega }$)
\begin{equation}
D(q,\omega )=b^{2-d}D(bq,\omega  b^d)
\end{equation}
which is required for a transformation of length scales $r\mapsto r/b$ 
\cite{AL86}. Note that in contrast to the conductance at $E_{c}$, the 
diffusion function itself is scale dependent. 

From these arguments it is not evident where the transition between
regions (B) and (C) occurs. It is conceivable that instead of the
scaling behavior in region (B) a smooth crossover from region (A) to
(C) takes place. From the present numerical results we cannot address
this question. 

The regions (A), (B), and (C) correspond to three different regimes of the function $f(x)$: First, in order to have $D(q\to 0,\omega )\sim 
L_\omega ^{2-d}$ (regime A), the function $f(x)$  must follow $f(x)\sim 
x^{(2-d)/d}$ in the limit $x \rightarrow 0$. Second, for large $x$, one 
has $f\sim x^{-\eta/d}$ in order to fulfill the behavior Eq.~(\ref{D}), 
regime C. The scaling form in the intermediate region (B),
$D(q,\omega )\sim q^{d-2}$, requires that $f(x)= \text{const}$ there. 
We thus can write the correlation function $S(q,E,\omega )$ as
\begin{equation}
S(q, E, \omega )=\frac{\rho(E)}{\pi\hbar\omega }
\frac{xf(x)}{1  +\left [ xf(x)\right]^2 },\qquad 
x=[q\cdot{\rm min}(\xi,L_{\omega })]^d,
\end{equation}
where
\begin{equation} 
\label{fx} 
f(x)= \left\{\begin{array}{r@{\quad:\quad}l} 
c_{\alpha }\cdot x^{(2-d)/d} & x \to 0\ \mbox{(A)}\\ 
c_{\beta} & \mbox{in between (B)} \\
c_{\gamma  }\cdot x^{-\eta/d} &  x\to \infty\ \mbox{(C)}
\end{array} \right.
\end{equation} 
with constants $c_{\alpha }$, $c_{\beta }$ and $c_{\gamma }$. In the brackets 
() we indicate the corresponding area in Fig.~\ref{eta3}. The boundaries 
between the three regions and the constants have to be extracted from 
numerical calculations. For the latter, we work close to $E_{c}$ so that
$\xi>L_{\omega }$ and the scaling variable $x=[q\cdot L_{\omega }]^d$.
The case $\xi<L_{\omega }$ gives rise to the two remaining areas in 
Fig.~\ref{eta3} with $D\sim \xi^{2-d}$ (left to A) and $D\sim q^{d-
2}(q\xi)^{-\eta}$ (left to C), which are not relevant in our analysis. 
In the  case of two dimensions, $d=2$, regions A and B coincide and $f$ is 
constant there. The numerical analysis (see below) shows that the
large-$x$ regime of $f(x)$, governed by the power law involving the
exponent $\eta$, is indeed relevant over several decades of $x$. 

We notice that upon Fourier transforming $S(q,\omega )$ into real space, 
large values of $x$ correspond to small $r\ll {\rm min}(\xi,L_{\omega })$, 
where the strong fluctuations of the eigenstate amplitudes become important. 
In this region (corresponding to $x=[{\rm min}(\xi,L_{\omega })/r]^d \gg 1$), 
one finds 
\begin{equation}
S(r,\omega )\sim \frac{1}{\omega r^{d}}x^{-1+\eta/d}\sim  
\frac{1}{\omega r^{d}} [{\rm min}(\xi,L_{\omega 
})/r]^{-d+\eta}
\end{equation}
which for $\omega \to 0$ ($\xi<L_{\omega }$) gives $S(r,\omega )\sim 
1/(\omega \xi^{d})\cdot (r/\xi)^{-\eta}$, Eq.~(\ref{Somega0}). For 
$\xi>L_{\omega }$ one recovers the form $S(r,\omega )\sim r^{-\eta}\omega ^{-
\eta/d}$ given by Wegner \cite{remarkwegner1}. 

\section{A Wave packet at the critical point}

In order to illustrate the implications of critical dynamics, we use the 
relation between the diffusion function $D(q,\omega )$ and the 
diffusion propagator $P(q,\omega )$, Eq.~(\ref{diffusionpropagator}).
The latter is the Fourier transform of the probability distribution 
$P({\bf r},t)$ describing the motion of an electron wave packet constructed 
from eigenstates with energies close to a fixed energy $E$. In the metallic 
region, $P({\bf r},t)$ is the solution of the diffusion equation in $d$ 
dimensions
\begin{equation}
\frac{\partial P({\bf r},t)}{\partial t}-D\nabla^{2}P({\bf r},t)=0,
\quad t>0
\end{equation}
with the diffusion constant $D$,
\begin{equation}
P({\bf r},t)=\frac{e^{-r^{2}/(4Dt)}}{(4\pi Dt)^{d/2}}, \quad t>0.
\end{equation}
Due to causality, $P(q,\omega )$ has to be analytical in the upper half plane 
as a function of complex $\omega $: The integral
\begin{equation}
P(q,t ) = \frac{1}{2\pi}\int d\omega \, \frac{e^{-i\omega t}}{-i\omega 
+D(q,\omega )q^{2}} 
\end{equation}
has to be performed in the upper half plane for $t<0$ whence $P(q,t<0)=0$. 
On the other hand, for $t>0$ we can write
\begin{equation}
P(q,t ) = \frac{1}{2\pi}\int d\omega \,e^{-i\omega t}\left\{
\frac{1}{-i\omega+D(q,\omega )q^{2}} +
\frac{1}{[-i\omega+D(q,\omega )q^{2}]^{*}} \right\},\quad t>0
\end{equation}
since for $t>0$ the integral is performed in the lower half plane
where the second term gives zero contribution. For real
$\omega $, $D(q,\omega )$ is real and one can write
\begin{equation}
P(q,t>0) = \frac{1}{\pi}\int d\omega \,e^{-i\omega t}
\frac{D(q,\omega )q^{2}}{\omega^{2}+(D(q,\omega )q^{2})^{2}}
= \frac{\hbar}{\rho }\int   d\omega e^{-i\omega t}S(q,E,\omega ).
\end{equation}
Recalling that $S(q,E,\omega )$ is an even function in $\omega $ and 
introducing the dimensionless variable $x=(qL_{\omega })^{d}\equiv 
q^{d}/\hbar\omega \rho $, we obtain
\begin{equation}
\label{Pqt}
P(q,t>0)=2\int_{0}^{\infty}d\omega \,\frac{\hbar}{\rho }\cos (\omega 
t)S(q,E,\omega )=\frac{2}{\pi}\int_{0}^{\infty}\frac{dx}{x}\,
\cos\left(\frac{q^{d}t}{\hbar\rho x}\right)
\frac{xf(x)}{1+[xf(x)]^{2}}.
\end{equation}
Normalization of the wave packet requires $P(q=0,t>0)=1$, i.e.
\begin{equation}
\frac{2}{\pi}\int_{0}^{\infty}\frac{dx}{x}\,
\frac{xf(x)}{1+[xf(x)]^{2}}=1
\end{equation}
which is an additional condition to be fulfilled by the function $f(x)$ 
Eq.~(\ref{fx}). Since the exact form of $f(x)$ is not known, we cannot 
determine $P(q,t)$ exactly. However, for large $q^{d}t$ 
the probability $P(q,t)$ in  Eq.~(\ref{Pqt}) 
is determined by large $x$ values due to the 
rapidly oscillating cosine-term. For large $x$, on the other hand, the 
form of $f(x)$ is known, and we can write
\begin{equation}
\label{Pqtasympt}
P(q,t)\sim \frac{2}{\pi}\int_{x_{-}}^{x_{+}}\frac{dx}{x}
\cos\left(\frac{q^{d}t}{\hbar\rho x}\right)
\frac{1}{c_{\gamma}x^{1-\eta/d}}\sim (q^{d}t)^{-1+\eta/d}=(q^{d}t)^{-D_{2}/d},
\quad q^{d}t/\hbar \rho \gg 1.
\end{equation}
Here, we used the relation $D_{2}=d-\eta$ (see the following section). 
Furthermore, the lower (high frequency) cutoff $x_{-}$ is determined by the 
inverse of a microscopic time scale $\tau $, below which the motion of the 
electron is ballistic and not described by the diffusion pole $P(q,\omega )$. 

In the same line of argument, one obtains the {\em probability of return} of 
a wave packet to the origin in the long time limit. This quantity is defined  
by the ${\bf r}=0$ value of $P({\bf r},t)$ and can therefore simply be 
recovered from $P(q,t)$ by Fourier transformation. Again, the upper limit of 
the resulting $q$-integral is determined by a microscopic cutoff $\sim 1/l$ .
One has
\begin{equation}
\label{wavepacketlong}
p(t)=\frac{1}{(2\pi)^{d}}\int dq^{d}\,P(q,t)\sim\int^{1/l}dq q^{d-1}\,
(q^{d}t)^{-D_{2}/d}\sim t^{-D_{2}/d}.
\end{equation}
This dependence of $p(t)$ is not valid for very short time scales 
corresponding to a (quasi)-ballistic motion of the wave packet, and times so 
long that the system boundaries become important.

We can compare the critical behavior of $p(t)$ in $d=2$ and $d=3$ to the 
metallic case, where $p(t)=(4\pi D t)^{-d/2}$:
\begin{equation}
p(t)\sim \left\{\begin{array}{r@{\quad:\quad}l} 
t^{-1}&\mbox{\rm 2d metallic}\\
t^{-(1-\eta/2)}&\mbox{\rm 2d critical}\\
t^{-3/2}&\mbox{\rm 3d metallic}\\
t^{-(1-\eta/3)}&\mbox{\rm 3d critical}\\
\end{array} \right.
\end{equation}
In particular, this demonstrates the different role of the exponent $\eta$ in
two and three dimensions: While in two dimensions, $\eta$ essentially can be 
regarded as a more or less small correction to the ordinary metallic 
diffusion pole, the situation in three dimensions is drastically different. 
There, due to the scaling form of $D(q,\omega )$, the dynamical behavior of 
electrons near $E_{c}$ is qualitatively very different from the metallic 
case.

\section{Relation between the correlation dimension $D_{2}$ and the 
exponent $\eta$} 

The relation Eq.~(\ref{relation}) has been derived previously 
\cite{BSK94c,Jan94}; we shortly outline our proof since it comprises
the  definitions concerning the multifractal dimension $D_{2}$ entering the 
analysis of the wave function at $E_{c}$. One first defines the probability 
\begin{equation}
P_{i}(\lambda ):=\int_{\Omega _{i}(\lambda )}d^dx \,|\Psi_{E}(x)|^2
\end{equation}
to find the electron of fixed energy $E$ in a finite region (box $i$) of real 
space denoted by $\Omega _{i}(\lambda )=l^{d}$ as a subspace of the 
total volume $\Omega =L^d $,  where $l=\lambda L, 0<\lambda \le 1$. 
The scaling with $\lambda $ is used to define so-called fractal dimensions
$D_{q}$. One considers the averaged $q$th power of $P_{i}(\lambda )$ :
\begin{equation}
\label{Plambda}
P(\lambda,q ):=\frac{1}{N(\lambda )}\sum_{i=1}^{N(\lambda )}\left [
P_{i}(\lambda)\right]^q\sim \lambda ^{D_{0}+(q-1)D_{q}}.
\end{equation}
Here, $N(\lambda )=\lambda^{-d}$ is the number of small cubes into which the original
total cube of volume $L^{d}$ was split. 
Requiring that at $E_{c}$, the summation in Eq.~(\ref{Plambda})
over the different boxes labeled by $i$ is equivalent to a disorder average 
in one (arbitrarily) fixed box (say $i=1$), 
\begin{equation}
P(\lambda,2 )=\left\langle P_{1}(\lambda)^{2}\right\rangle,
\end{equation}
one has (omitting the index $i=1$ now)
\begin{equation}
P(\lambda,2 )=\int_{\Omega (\lambda )}d^d x\,\int _{\Omega (\lambda )} d^d 
x'\,\left\langle|\Psi_{E}(x)|^2|\Psi_{E}(x')|^2\right\rangle.
\end{equation}
Because of the disorder average, the product of the squares should depend 
on the difference $x-x'$ only. One can then introduce coordinates $r=x-x'$ and 
$R=(x+x')/2$. The Jacobian of this 
transformation is 1, but because the integration region is finite, the 
integration limits are also changed which makes the evaluation tedious.
However, for large enough $\Omega (\lambda )$ the integral
over $R$ simply gives the volume $\Omega (\lambda )$ itself, and we have
\begin{equation}
P(\lambda,2 )=\Omega (\lambda ) \cdot \int _{\Omega (\lambda )} d^d 
r\,\left\langle|\Psi_{E}(0)|^2|\Psi_{E}(r)|^2\right\rangle. 
\end{equation} 
Considering the definition of the two-particle correlation function
Eq.~(\ref{Sdefinition}) for $\omega \to 0$, one obtains 
\begin{equation}
P(\lambda,2 )=\left(\Omega \rho(E)\right)^2\Omega (\lambda )\cdot  
\int _{\Omega (\lambda )} d^d r\, S(r,E,\omega \to 0).
\end{equation}
where $\Omega $ is the total volume $\Omega =\Omega (\lambda =1)$ of the 
system.
Using Eq.~(\ref{Somega0}), one has
\begin{equation}
\int_{\Omega (\lambda )} d^d rr^{-\eta}\sim
\int_{0}^{\lambda L} d r r^{d-1} r^{-\eta}\sim
\left(\lambda L\right)^{d-\eta}.
\end{equation}
Because $\Omega (\lambda )=(\lambda l)^d$, we find
\begin{equation}
P(\lambda,2 )\sim \Omega (\lambda )\cdot \int _{\Omega _{\lambda }}d^dr r^{-
\eta}\sim \lambda ^d \lambda ^{d-\eta}.
\end{equation} 
By comparison with Eq.~(\ref{Plambda}) one reads off
$ d + (d-\eta) = D_{0}+D_{2}$.
Noticing that $D_{0}=d$ is the dimension of the total support of the 
wave function, we have the announced relation Eq.~(\ref{relation}) between the 
exponent 
$\eta$ and the fractal dimension
\begin{equation}
D_{2}=d-\eta.
\end{equation}
This fundamental relation relates 
properties of correlation functions to the spatial structure of the 
wave functions at the critical point. Furthermore, if $\eta$ is
larger than $0$ at a critical point of a metal-insulator transition
\cite{remarkwegner2}, by Eq.~(\ref{relation}) it follows $D_{2}<d$ and the 
wave functions must be multifractal \cite{Weg80}. 

We obtain another exponent relation by exploiting 
the long-time behavior $p(t)$ of the return probability of a wave packet, constructed from eigenstates near $E_{c}$. 
First, the exponent $\delta$ introduced via $p(t)\sim t^{-\delta}$ 
was generally proven by Ketzmerick et al. \cite{KPG92} to be equal to
the generalized dimension $\widetilde{D_{2}}$ of the spectral 
measure. Just as $D_2$ describes spatial correlations,
$\widetilde{D_{2}}$ describes the correlations of the {\em local}
density of states of the system as a function of energy. In 
\cite{KPG92} it was shown that $\delta=\widetilde{D_{2}}$. On the other 
hand, we have seen that $\delta =D_{2}/d$, Eq.~(\ref{wavepacketlong}), and 
therefore      
\begin{equation}
\delta=\frac{D_{2}}{d}=\widetilde{D_{2}}.
\end{equation}
     
\section{Numerical Analysis}

\subsection{Model for the numerical investigation}

The dynamics of non interacting electrons in the presence of disorder is 
studied within the framework of the Anderson model described by the 
Hamiltonian
\begin{equation}
{\cal H} = \sum_{\bf r} \epsilon_{\bf r} |{\bf r}\rangle\langle{\bf r}|
+ \sum_{\langle\bf r,r'\rangle} V_{{\bf r,r'}} |{\bf r}\rangle\langle{\bf r'}|,
\end{equation}
where the diagonal disorder potentials $\epsilon_{\bf r}$ are independent 
random numbers with a constant probability distribution in the range 
$-W/2 \le \epsilon_{\bf r} \le W/2$ and the non diagonal transfer matrix 
elements between nearest neighbors, $V_{\bf r,r'}$, are taken to be the unit 
of energy. The vectors ${\bf r}$ denote the sites of a simple cubic lattice 
with lattice constant $a$ and periodic boundary conditions are applied in all 
directions.

Eigenvalues and eigenvectors have been obtained for systems of size up to 
$(L/a=40)^3$
sites by direct diagonalization using a Lanczos algorithm. 
The critical regime in the middle of the disorder broadened tight binding band,
where both the localization length and the correlation length diverge, is known
to correspond to a critical disorder of $W_c/V\simeq 16.4$ \cite{HS93,ZK95} 
which separates localized ($W > W_c$) from metallic behavior ($W < W_c$).

\subsection{The function $Z(\omega )$ and the exponent $\mu=\eta/d$}

For the numerical analysis, we defined the function
\begin{equation}
Z(E,E'):=\int d^d x\,|\Psi_{E}(x)|^2 |\Psi_{E'}(x)|^2
\end{equation}
for eigenstates with energy $E$ and $E'$, where $d$ denotes the spatial 
dimension. 
In the following, we will be interested in the case where the localization 
length is the largest length scale in the system so that $\xi(E)>L_{\omega }$. 
We show that one can directly extract the quantity $\eta$ from 
$Z(E,E')$ which is easier to obtain numerically than, e.g., a direct 
determination from the diffusion function $D(q,\omega )$ itself. 
Indeed, the two-particle correlation function in real space for ${\bf r}=0$,
\begin{equation}
\label{sr0}
S(r=0,E,\omega):= \left\langle
\sum_{\alpha ,\beta }\delta (E^- -E_{\alpha }) \delta (E^+ -E_{\beta})
\Psi_{E^-}(0) \Psi_{E^-}^*(0)\Psi_{E^+}(0)\Psi_{E^+}^*(0)\right\rangle
\end{equation}
can be related to $Z(E,E')$ assuming that the disorder average in 
Eq.~(\ref{sr0}) is equivalent to a {\em spatial} average for one 
fixed impurity configuration. In this case,
\begin{equation}
\label{SZ}
S(r=0,E,\omega )=\Omega ^{-1}\left(\Omega\rho(E^-)\right)
\left(\Omega\rho(E^+)\right) \int d^dx\,  |\Psi_{E^-}(x)|^2 |\Psi_{E^+}(x)|^2,
\end{equation}
with $\hbar\omega=E^+ - E^-$ and $E=(E^+ + E^-)/2$. 
The equivalence of these averages is underlying the subsequent 
analysis of the function $Z(E,E')$.
The relation
\begin{equation}
\label{Z}
Z(E^+,E^-) = \frac{1}{(2\pi)^d \Omega \rho(E^+)\rho(E^-)}\int d^dq\,
S(q,E,E^+ -E^-),
\end{equation}
obtained by Fourier transformation, is used in the following. 
Keeping $E$ fixed at the critical energy, the quantity $Z$ depends on $\omega 
=(E^+ -E^-)/\hbar$ only.
Since the numerical calculation is performed on a finite lattice, 
the $q$-integrals have to be cut off at
$q=2\pi/a$, where $a$ is the lattice constant. A lower cutoff is given 
by the system size $L$ itself. In terms of the scaling variable $x$, the lower 
cutoff is $x_{-}(\omega):=((L/2\pi)^d\rho\hbar\omega)^{-1}=(2\pi)^d
\Delta/\hbar\omega$.   
The upper cutoff is $x_{+}(\omega ):=((a/2\pi)^d\rho\hbar\omega)^{-1}
= x_{-}(\omega )(L/a)^d$. In a finite system, energy 
transfers $\hbar\omega $ from one state with energy $E$ to another with 
energy $E+\hbar\omega$ satisfy $\hbar\omega \gtrsim \Delta $. In terms
of the variable $x$, this means
$  x_{-}(\omega)\lesssim (2\pi)^{d}$. For the function $Z(\omega )$, 
$\hbar\omega =E^+-E^-$ in
Eq.~(\ref{Z}), and we obtain
\begin{equation}
\label{Zintegral}
Z(\omega )=\frac{s_{d}}{(2\pi)^{d}\Omega \rho ^{2}} \int_{2\pi/L}^{2\pi/a}dq\,
q^{d}      
S(q, E, \omega )=\frac{s_{d}}{d\pi(2\pi)^{d}\Omega } 
\int_{x_{-}(\omega)}^{x_{+}(\omega)}dx\,\frac{xf(x)}{1+[xf(x)]^2}.
\end{equation} 
Here, $s_{d=2}=2\pi$ and $s_{d=3}=4\pi$. 
For small $\omega $ and large $L$ the integral is dominated by the
large $x$ behavior of $f(x)$ (region (C)). 
\begin{equation}
\label{zomega}
Z(\omega )\approx
\frac{s_{d}}{d\pi(2\pi)^{d}\Omega }\int_{x_{-}(\omega)}^{x_{+}(\omega )} 
dx\,(x c_\gamma x^{-\eta/d})^{-1}\approx 
\frac{s_d}{\pi(2\pi)^d\Omega c_\gamma \eta} \left(\frac{2\pi
    L}{a}\right)^\eta \left(\frac{\Delta}{\hbar\omega}\right)^{\eta/d}.
\end{equation} 
For small $\omega $, there is a range where $Z(\omega )\sim \omega
^{-\mu}$  with $\mu=\eta/d$. 
A logarithmic plot of $Z(\omega )$ thus yields the exponent $\mu=\eta/d$, 
furthermore from Eq.~(\ref{zomega}) one can determine the coefficient
$c_{\gamma }$. 

\subsubsection{Numerical data}

\begin{figure}[htb]
\unitlength1.0cm
\begin{picture}(12,10)
\put(-1,-18){\epsfbox{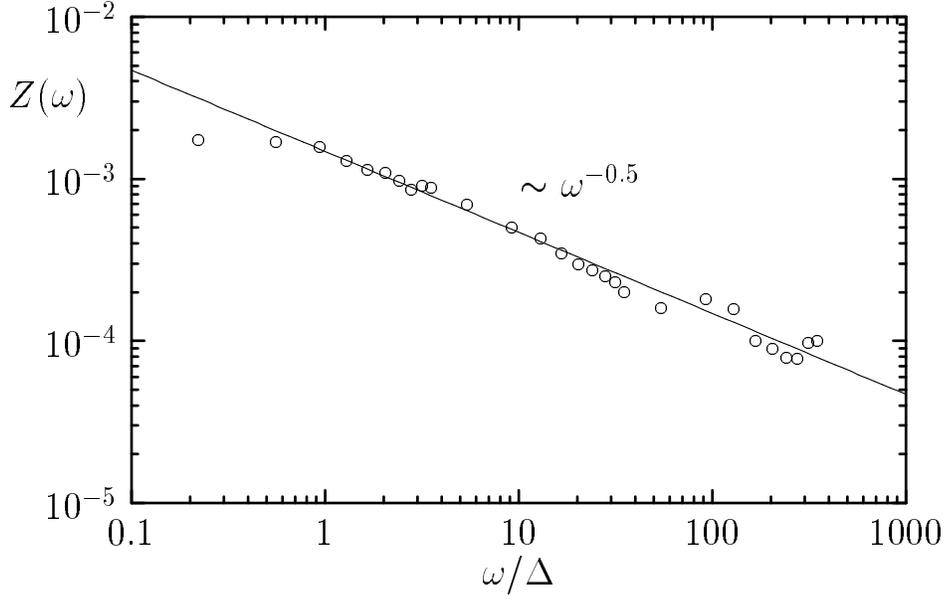}}
\caption[]{\label{zomdd} Energy correlation function $Z(\omega)$ of 
eigenstates taken 
from the critical regime ($W_c/V=16.3$) of a 3d Anderson model of system size 
$(L/a=40)^3$. We used 371 eigenstates with energies $E$ from the interval 
$-0.05\le E/V \le 0.05$ to extract the exponent $\mu =\eta/3=0.5$ from the 
power law decay.}
\end{picture}
\end{figure}

We used the numerical data to obtain the exponent $\mu=\eta/d$ from the 
function $Z(\omega )$. The result for a system of size $\Omega =(40\,a)^{3}$ 
is shown in Fig.~\ref{zomdd}.  
One clearly observes a power law with an exponent $\mu=0.5\pm 0.1$. 
This yields $\eta=\mu\cdot d\approx 1.5$. In the energy range $\omega_c 
\approx 300\Delta >|E-E'|\gtrsim \Delta$, we can fit the function $Z$ by 
$Z(\omega )=Z_{0}|E-E'|^{-0.5}$ with $Z_{0}=2.5\cdot 10^{-5}$. 
$\Delta/V$ is about $2.7\cdot 10^{-4}$ in the present system. 
` For even larger values of $\omega$ or
correspondingly smaller $x$, one should enter regions (A) and (B)
which are beyond the applicability of eq.~(\ref{zomega}).

On the other hand, for small $\omega \lesssim \Delta$, $Z(\omega)$
saturates at the value of the inverse participation ratio $P^{(2)}(E)=\int
d^d x|\Psi_E(x)|^4\propto (L/a)^{-D_2}$. A similar limiting behavior has also
been observed in the QHE-case (see below) and in the 2d-symplectic situation 
\cite{Sch95}. 

\subsubsection{Comparison to the quantum Hall case ($d=2$)}

\begin{figure}[]
\unitlength1.0cm
\begin{picture}(12,10)
\put(-1,-18){\epsfbox{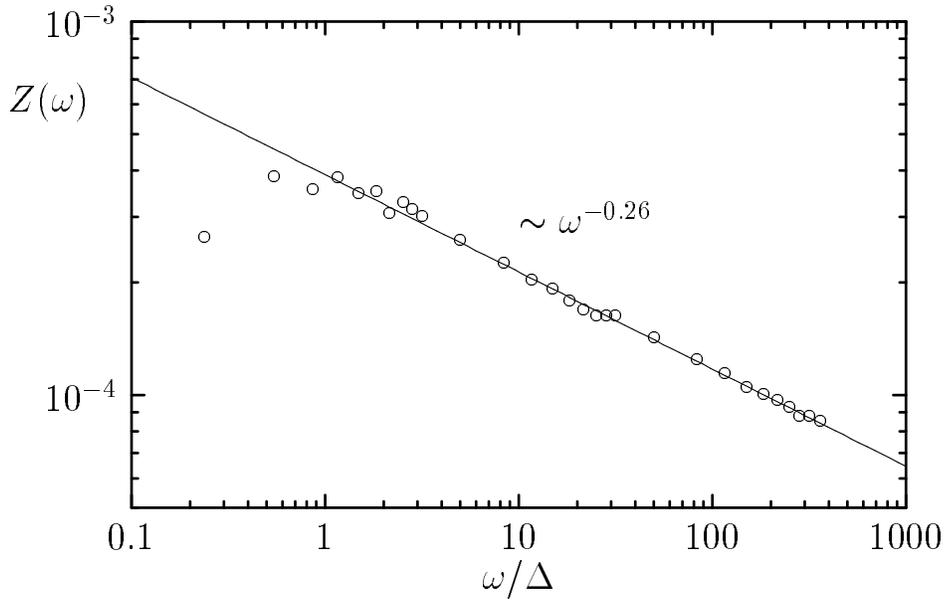}}
\caption[]{\label{zomqhe}Energy correlation function $Z(\omega)$ for a 
QHE-system of 
size $(L/a=125)^2$. 330 eigenstates taken from an energy interval 
$\Delta E=0.1\,V$ around the center of the lowest Landau
band were used to determine the exponent $\mu=\eta/2=0.26$.} 
\end{picture}
\end{figure}

In the quantum Hall case there exists no complete Anderson transition, but
for finite systems a critical behavior and multifractal eigenstates 
\cite{PJ91,HS92,HKS92} can be 
observed in an energy range about the center of the disorder broadened Landau 
band, $E_0$, where the localization length exceeds the system size,
$\xi=\xi_0|E-E_0|^{-\nu} >> L$. We used our previously obtained eigenvalues
and eigenfunctions of a system of size $125\times 125$ and a magnetic field
which corresponds to $1/5$ flux quanta per plaquette (see \cite{HS94} for 
details) to calculate the function $Z(\omega)$ shown in Fig.~\ref{zomqhe}. 
Again, a power law relation can be observed with a saturation at $\omega 
\approx \Delta = 3\cdot 10^{-4}\,V$. The exponent $\mu=\eta/2=0.26\pm 0.05$ 
is somewhat larger than the value obtained from the exponent $\delta$ of the
temporal decay of a wavepacket ($\eta/2 = 1-\delta=0.19$) built from the same 
critical eigenvectors \cite{HS94}.

\subsection{Exponents $D_{2}$, $\widetilde{D_{2}}$, $\delta $.}
\subsubsection{The Wave packet}

\begin{figure}[]
\unitlength1.0cm
\begin{picture}(12,10)
\put(-1,-18){\epsfbox{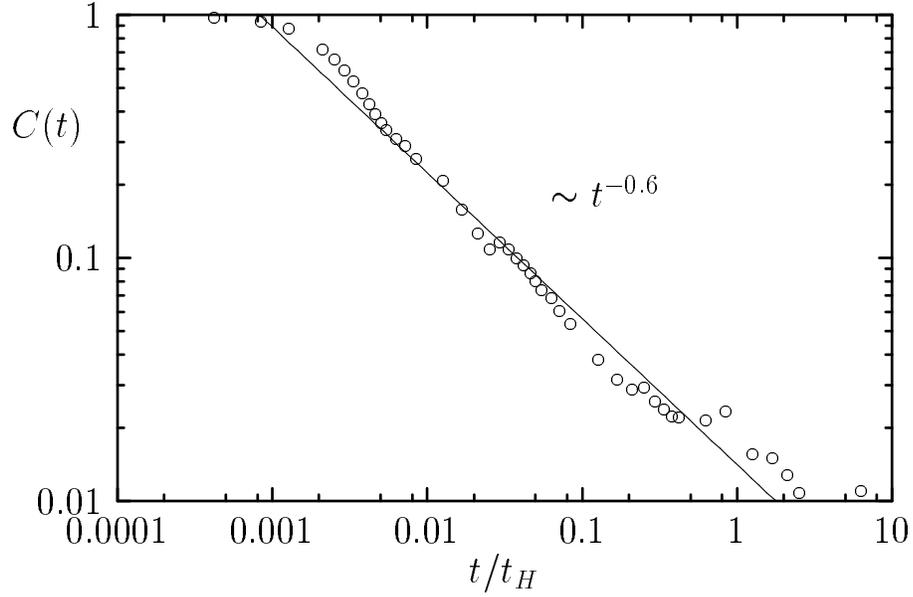}}
\caption[]{\label{coft3d}Temporal autocorrelation function $C(t)$ for a 3d
critical Anderson model ($W_c/V=16.5$) of size $(L/a=35)^3$ versus time $t$
in units of the Heisenberg time $t_H=\hbar/\Delta$. To construct the
wave packet at the initial site a total of 1194 eigenstates from the energy
interval [$-0.5,0.0$] have been used.}
\end{picture}
\end{figure}

The correlation function $C(t)$ related to the probability of return $p(t)$
\cite{KPG92,HS94} of a wave packet, constructed from eigenstates with energies
$E\approx E_{c}$ of a 3d Anderson model, is shown in
Fig.~\ref{coft3d}, where $C(t)$ is plotted as function of $t\Delta/\hbar$.
A power law $p(t)\sim t^{-\delta}$ can be clearly identified, we
obtain $\delta = 0.6\pm 0.05$ in the region where $t > 2\cdot
10^{-3}\hbar/\Delta$. The condition
$q^{d}t/\hbar\rho $ in Eq.~(\ref{Pqtasympt}) can be checked if we use
$q=2\pi/L$ as the smallest $q$ value in the system ($L/a=35$). Together with
the density of states, which is  $\rho =0.058/(Va^3)$, we obtain
from Eq.~(\ref{Pqtasympt}) the condition $t\gg 4\cdot
10^{-3}\hbar/\Delta$ which is in accordance
with the observed behavior in Fig.~\ref{coft3d}.

\subsubsection{Multifractality exponents}

\begin{figure}[]
\unitlength1.0cm
\begin{picture}(12,10)
\put(-1,-18){\epsfbox{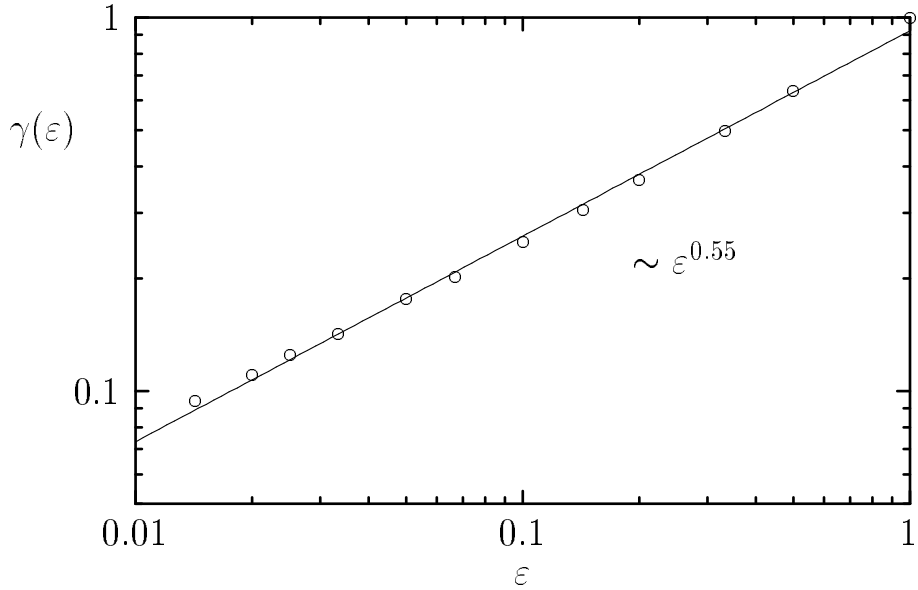}}
\caption[]{\label{gamma2_3d}The correlation dimension $\widetilde{D_2} = 0.55
\pm0.5$ of the spectral measure as obtained from the scaling relation of 
$\gamma_2$ (see Eq.~\ref{gamma_q}) versus the length of the energy interval 
$\varepsilon$ for a critical system of size $(L/a=40)^3$.}
\end{picture}
\end{figure}

We calculated the correlation dimension of the multifractal wavefunction
$D_{2}$, Eq~(\ref{Plambda}), and the correlation dimension of the spectral 
measure, $\widetilde{D_{2}}$, defined by
\begin{equation}
\gamma(q,\varepsilon )=\lim_{\varepsilon \to 0}1/L^{2}\sum_{{\bf r}} \sum_{i}
\Big(\sum_{E \in \Omega _{i}(\varepsilon) } |a_{E}|^{2}\Big)^{q} 
\sim \varepsilon ^{(q-1)\widetilde{D}_{q}},
\label{gamma_q}
\end{equation}
using a box-counting method.
The result of the scaling behavior of the latter is shown in 
Fig.~\ref{gamma2_3d}. 
The exponent $\delta $ from the probability of return of a wave packet 
(Fig.~\ref{coft3d}) and the most probable scaling exponent, $\alpha _{0}$, 
determining the characteristic $f(\alpha)$-distribution (see e.g. \cite{Jan94}
and references therein) related to the generalized fractal dimensions $D_q$ 
which completely describe all the moments of the spatial amplitude 
fluctuations of the critical wave functions have also been calculated. 
Our results are compiled in Table 1. 
The relations concerning the exponent $\eta$ which have 
to be fulfilled according to Eq.~(\ref{wavepacketlong}), Eq.~(\ref{relation}),
and the definition $\mu=\eta/d$, are $\eta=d(1-\delta)=1.2\pm 0.15$, 
$\eta=d-D_{2}=1.3\pm 0.2$, and $\eta=1.5\pm0.3$, respectively. 
Thus, one can say that within the numerical uncertainty the different methods 
of determining $\eta$ yield the same value $\eta\approx 1.3$. 
We note that our value for $D_{2}$ is very close to a result of previous 
numerical work by Soukoulis and Economou using a different method
\cite{SE84} who obtained $D_{2}= 1.7\pm 0.3$. 

\section{Inelastic Scattering at the Anderson-Transition}

In this section we investigate the implications of critical dynamics 
at the Anderson transition on inelastic scattering rates. 
Inelastic scattering rates were calculated previously 
\cite{previously1} for a quantum Hall 
system where the critical energy coincides with the Landau band center. It 
could be shown that the exponent $\eta$ describing the eigenfunction 
correlation showed up in the temperature dependence of the scattering rate 
and the energy loss rate of electrons with acoustical phonons. On the other 
hand, the temperature dependence of the electron-electron scattering rate 
was similar to that in a two-dimensional disordered metal without a magnetic 
field. It is therefore of some interest to study these quantities for the 
three-dimensional system, too. 

\subsection{Electron-phonon scattering rate}

The electron-phonon (e-p) scattering rate is calculated 
in the standard way \cite{Mahan} to second order in $V_{q}$, the e-p 
coupling matrix element. It is assumed that the impurity potential is
not changed by the lattice motion. In a metal this assumption is not 
fulfilled a priori: The impurities are embedded in the lattice and 
therefore move in phase with the other lattice atoms \cite{CS86}. Here, we 
address to a situation where the random potential is generated by external 
sources like donor atoms far away from the electron gas.

The imaginary part of the self energy gives the 
inelastic lifetime of an electron of energy $E$
\begin{eqnarray} \label{tau-1}
 \lefteqn{ \tau ^{-1}_{ep}(E) = \frac{2\pi}{\hbar\Omega}\sum_{{\bf q},E'} 
|V_{q}|^{2}|M_{E,E'}({\bf q})|^{2}
 \times} \\ \nonumber
 &\times&\left\{[n(\omega _{q})+f(E')]\delta(E -E'
+\hbar\omega_{q}) - (\omega_{q}\rightarrow -\omega_{q}) \right\}
\end{eqnarray}
Here, $n(\omega_{q})$ denotes the Bose function 
for phonon frequency $\omega _{q}$ and $f$ the Fermi function.
The terms in the curly brackets correspond to phonon emission 
($+\hbar\omega _{q}$) and -absorption ($-\hbar\omega _{q}$), respectively. 
The $E'$-sum runs over all intermediate eigenstates of energy $E'$ which 
appear in the momentum matrix element $M_{E,E'}({\bf q})=
\langle E |\exp (i{\bf q}{\bf x})|E'\rangle$.
The central quantity containing the information about the unperturbed system 
in (\ref{tau-1}) is  
\begin{equation}
\sum_{E'}|M_{E,E'}({\bf q})|^{2}\delta(E -E'
+\hbar\omega_{q})=\Omega\rho(E+\hbar\omega _{q})
|M_{E,E+\hbar\omega _{q}}({\bf q})|^{2},
\end{equation}
where $\rho$ is the three-dimensional density of states and 
$\Omega$ is the total volume of the system.
We express the matrix elements by the two-particle correlation function
\begin{equation}
\Omega\rho(E+\hbar\omega) |M_{E,E+\hbar\omega}({\bf q})|^{2}
= \frac{1}{\rho(E)}S(q;E+\frac{\hbar\omega }{2},\omega )
\end{equation}
We restrict the discussion to energies
$E=E_{F}$, where $E_{F}$ is the Fermi energy.
Using $n(\omega _{q})+f(E+\hbar\omega _{q})=1/\sinh (\beta\hbar\omega _{q})$,
we get
\begin{equation}
\label{tauep}
\tau^{-1}_{ep}=\frac{4}{2\pi\hbar}\int_{0}^{\omega_{D}} d\omega \,
\frac{\alpha ^2\! F(\omega )}{\sinh \beta \hbar \omega }
\frac{S(\frac{\omega }{c_{S}},E+\frac{\hbar\omega }{2},\omega)}{\rho(E)}
\end{equation}
where we defined the function $\alpha ^{2} \! F(\omega )=\alpha ^{2}\omega 
^{n}$
for acoustical phonons with coupling constant $\alpha ^{2}$. 
In the case of non-piezoelectric materials, only the 
deformation potential coupling is relevant. Then, $n=3$ and $\alpha ^{2}= 
(\Xi^{2}\hbar)/(2\rho_{M}c_S^{5})$ with speed of sound $c_{S}$, 
deformation potential $\Xi$, and mass density $\rho_{M}$. 
The temperature dependence at the critical energy $E=E_{c}$ is determined by 
the form of the correlation function, defined via Eq.~(\ref{fx}) with $d=3$.
The phonon dispersion $\omega=c_{S}q$ leads to $x=\omega 
^2/(c_{S}^3\rho\hbar)$. The dimensionless variable $x$ ($d=3$) can be written 
as $x=(\omega/\omega_{cr})^2$ by introducing the crossover frequency 
\begin{eqnarray}
\omega _{cr}:=\left(\hbar c_{S}^3\rho\right)^{1/2}.
\end{eqnarray}
The exponent for the $T$-dependence of $\tau ^{-1}_{ep}$ can be extracted in 
two limits:
For frequencies $\omega \gg\omega _{cr}$ one has $x\gg 1$ and the correlation 
function
$S\sim (xf(x))^{-1}/\omega  \sim \omega ^{-3+2\eta/3}$. In the opposite case 
$\omega \ll\omega _{cr}$ and $x\ll 1$, $S\sim (xf(x))/\omega 
\sim \omega ^{4/3-1}$. This leads to different behaviors of the scattering 
rate $\tau_{ep}^{-1}$ in the limit $k_{B}T \ll \hbar\omega_{cr}$ and $k_{B}T 
\gg \hbar\omega _{cr}$, respectively.
We can evaluate Eq.~(\ref{tauep}), introducing  
$\beta\hbar\omega $ as new variable. First,
for temperatures $k_{B}T\gg\hbar\omega 
_{D}\gg\hbar\omega _{cr}$ larger than the Debye energy, the dependence is 
linear in $T$. This is (as in an ordinary metal) the trivial high-temperature 
case.
The more interesting limit $k_{B}T \ll\hbar\omega_{D} $ gives, 
together with $\alpha ^{2} \! F(\omega )\sim \omega ^n$ and the knowledge of 
the correlation function $S$, the limiting forms 
\begin{equation} 
\tau_{ep}^{-1}
\sim\left\{\begin{array}{r@{\quad:\quad}l} T^{n+4/3} & k_{B}T \ll 
\hbar\omega_{cr}\\ T^{n-2+2\eta/3} & k_{B}T \gg \hbar\omega_{cr}. 
\end{array}\right.
\end{equation} 
These two regimes correspond to the case (A) and (C) in Eq.~(\ref{fx}). The 
full temperature dependence of $\tau ^{-1}_{ep}$ can be obtained from a 
numerical evaluation of Eq.~(\ref{tauep}).

The result $\tau_{ep}^{-1}\sim T^{3+4/3}$ for deformation potential 
scattering ($n=3$) in the low temperature limit is a direct consequence of 
the 'Wegner-scaling' of the diffusion function $D(q,\omega )\sim \omega 
^{1/3}$ in regime (A). It should be compared to the dependence $\tau_{ep}^{-
1}\sim T^4$ for the corresponding rate in the disordered metallic case 
\cite{CS86}. We also note that as in the quantum Hall case 
\cite{previously1}, the exponent $\eta$ appears only in an intermediate and 
not in the low-temperature regime.

\subsection{Electron-electron scattering}

Electron-electron (e-e) scattering rates at the Anderson transition have 
been calculated first by Belitz and Wysokinski \cite{BW87}, who used the 
so-called 'exact eigenstate formalism' and an RPA-like approximation for 
the Coulomb interaction. The advantage of this method is that it is 
non-perturbative in the disorder and, at least in principle, applicable to
a large number of systems. By this one can generalize results for e-e 
scattering rates which are already known in the ballistic or the diffusive, 
weakly localized case. However, one has to keep in mind that this approach is 
perturbative in the interaction between the electrons.

The general expression for the electron scattering rate $\tau_{e}^{-1}$
at finite temperature $k_{B}T = 1/\beta >0$,
given by the on-shell electronic self-energy at Fermi energy $E=E_{F}$
\cite{AALR81,BW87}:
\begin{eqnarray}
\label{taue}
\tau_{e}^{-1} &=& 2\int_{0}^{\infty}d\omega \, \frac{\phi(\omega )}{\sinh 
\beta \hbar \omega } \nonumber \\
\phi(\omega ) &=& -\frac{1}{\pi\hbar \rho \Omega}\sum_{{\bf q}}{\rm Im}\,V(q,
\omega 
){\rm Im}\, \Phi(q,\omega )
\end{eqnarray}
can be derived by starting from the exact eigenstates of the unperturbed 
system \cite{AALR81}. Here, $\Phi(q,\omega )$ is the density relaxation 
function of which the imaginary part is related to that of the density 
response function $\chi(q,\omega )$ by ${\rm Im}\Phi(q,\omega ) =(1/\omega) 
{\rm Im}\chi(q,\omega ) = \pi S(q;E,\omega )$ where the energy $E$ appears in 
the  density of states $\rho =\rho (E)$. Notice that ${\rm Im}\Phi(q,\omega 
)={\rm Im}\Phi(q,-\omega ) $ and ${\rm Im}V(q,\omega )=-{\rm Im}V(q,-\omega)$ 
because ${\rm Re}\chi(q,\omega )$ is an even and ${\rm Im}\chi(q,\omega )$ is 
an odd function of $\omega $. One therefore can restrict the integration 
to positive values of $\omega$.  
The above formula can be viewed as the rate for the scattering of {\em one}
electron at the fluctuations of the electromagnetic field caused by 
the motion of all the other electrons \cite{Eil84},
the function $\phi$ being an effective density of states
of these  fluctuations. It corresponds in the case of electron-phonon
scattering to the Eliashberg-function $\alpha^2\!F$ which there 
essentially gives the phonon density of states.
In the Coulomb potential, screening is included via the RPA approximation
\begin{equation}
V(q,\omega )=\frac{V_{0}(q)}{1+\chi(q,\omega )V_{0}(q)}
\end{equation}
with the bare Coulomb potential $V_{0}(q)=4\pi e^{2}/q^{2}$.
Notice that only the inclusion of dynamical screening leads 
to an non vanishing imaginary part of $V(q,\omega )$ and thus to a nonzero e-e 
scattering rate. The bare Coulomb potential would render the electron 
self-energy merely real and $\tau_{ee}^{-1}$ would be zero.
The influence of arbitrary disorder on the screening here is incorporated in 
the correlator $\chi$ which contains all the information about the 
unperturbed system.
In \cite{BW87}, the $q=0$ limit of the diffusion function $D(q=0,\omega )$
was used for the evaluation of Eq.~(\ref{taue}) at the critical point. 
Although the integration requires both the $q$- and the 
$\omega $-dependence of the diffusion function, it turned out that this 
approximation affects only the prefactor of the scattering rate and not its 
temperature dependence.

In the following, we will use the full form of the diffusion function as it 
enters into the density-density correlation function, which at the critical 
point can be written as 
\begin{eqnarray}
\chi(q,\omega )=i\rho\frac{xf(x)}{1+ixf(x)},\quad x=\frac{q^3}{\hbar\omega 
\rho},
\end{eqnarray}
The function $\Phi(\omega )$ can than be evaluated; its 
zero-frequency limit is given by 
\begin{eqnarray}
\label{phi0}
\Phi(\omega 
=0)\equiv\Phi_{0}=\frac{1}{6\pi^3}\int_{0}^{\infty}dx\,\frac{1}{1+(xf(x))^2}.
\end{eqnarray}
As in the two-dimensional diffusive and Quantum Hall case 
\cite{Sia91,previously1}, the
dimensionless quantity $\Phi_{0}$ can be expressed by an integral over a 
function of the variable $x=q^d/(\hbar\omega \rho)$ where $d$ is the 
dimension. As in the above cases, the scattering rate
according to Eq.~(\ref{taue}) formally diverges because the function 
$\Phi(\omega )$ describing the DOS of the electromagnetic fluctuations from 
the electron 'bath' does not vanish at zero frequency. 
However, for finite temperatures such that
the  scattering rate $\tau^{-1}_{e}$ is larger than the frequency 
$\omega $, the description of the critical dynamics in terms of 
$\chi(q^3/\omega )$ can no longer be valid and the integral should 
be cut off at $\tau_{e}^{-1}$ \cite{BW87}. 
This leads to a self-consistent 
equation (for low $T$, $\Phi(\omega )$ can be replaced by $\Phi_{0}$)
\begin{equation}
\tau_{e}^{-1} = 2\Phi_{0}\int_{\tau_{e}^{-1}}^{\infty}d\omega 
\,\frac{1}{\sinh \beta \hbar\omega },
\end{equation}
of which the solution can be written as
\begin{equation}
\label{taueresult}
\tau_{e}^{-1}=\frac{k_{B}T}{\hbar}\gamma,\qquad 
\gamma =-2\Phi_{0}\ln\tanh(\gamma/2).
\end{equation}
This is in fact the result \cite{BW87} of Belitz and Wysokinski, differing
only in the prefactor given by
\begin{equation}
\gamma =-\frac{\sqrt{3}}{4c}\ln\tanh(\gamma /2)\qquad 
{\rm Ref.~\cite{BW87}},
\end{equation}
where $c=13.4$ \cite{AL86} or 1 \cite{BW87}, respectively. 
In view of our analysis, Eq.~(\ref{fx}) and Eq.~(\ref{phi0}), we find
that the prefactor depends on the microscopic details of the system,
namely the constants $c_{\alpha }$, $c_{\gamma }$, and the
dimensionless conductance at the critical point $g^{*}$. In
two-dimensional quantum Hall systems the critical conductance was
found to be universal \cite{HHB93}. However, we know of no argument
that shows the critical conductance to be universal in three
dimensions. Thus we find that the universality of the
rate $\tau _{e}^{-1}$ claimed in \cite{BW87} is related to the
universality or lack thereof of the critical conductance $g^{*}$
\cite{eeremark}.

\section{Conclusion}

We have discussed some aspects of multifractal fluctuations near the
Anderson transition in three-dimensional electron systems. The
behavior of the dynamical diffusion function $D(q,\omega)$ was
reviewed. For large wavevectors $q$ and small frequencies $\omega$
multifractal fluctuations lead to the occurrence of an anomalous
diffusion exponent $\eta$ in the diffusion function. We have found
$\eta\approx 1.3$ in a numerical analysis of eigenfunctions obtained
by numerical diagonalization, and verified that $\eta$ is related to
the correlation dimension $D_2$ of the multifractal fluctuations by
$\eta=d-D_2$.

The multifractal eigenfunction fluctuations influence the temperature
dependence of the electron-phonon and electron-electron scattering
rates. We found that the low-temperature exponent of the former is
modified by the finite value of $\eta$, while the latter is linear in
$T$ and depends on the dimensionless conductance.

\section{Acknowledgments}
T.B. would like to acknowledge support by the EU STF9 fellowship and
stimulating discussions with A. Kawabata and Y. Hirayama.
B.H. would like to acknowledge the support through the
Sonderforschungsbereich 341 of the Deutsche Forschungsgemeinschaft and
fruitful discussions with M. Jan{\ss}en.


\begin{thebibliography}{10}
\bibitem{remarkwegner1}
{The case $\xi>L_{\omega }$ with $S(r,\omega )\sim r^{-\eta}\omega ^{-\eta/d}$
  is identical to Eq. (5.15) in \cite{Weg76}}, where the exponent $\kappa '$
  corresponds to $\eta$.
\bibitem{remarkwegner2}
{In \cite{Weg76}}, $0\le\eta\le d$ is shown by means of Schwarz's inequality.
  Note that the exponent $\kappa'$ in \cite{Weg76} is $\eta$ in our notation,
  furthermore $\kappa'=-x_{2s}$ in \cite{Weg80}, cp. Eq. (3.21) and (1.6)
  there.
\bibitem{eeremark}
{For a quantitative estimate of $\gamma $}, one has to know the full form of
  $f(x)$ which, in principle, can be extracted from the numerical calculation.
  However, we find no direct relation of the constants $c_{\alpha }$ and
  $c_{\gamma }$ to an experimentally relevant quantity like the dimensionless
  conductance $g^{*}$.
\end{thebibliography}

\tabcolsep4mm
\begin{table}
\begin{center}
\begin{tabular}{c|c|c|c|c|c|c} 
\multicolumn{1}{l}{System} & \multicolumn{1}{|c}{$\nu$} & 
\multicolumn{1}{|c}{$\delta$} & 
\multicolumn{1}{|c}{$\alpha _{0}$} & \multicolumn{1}{|c}{$D_{2}$} & 
\multicolumn{1}{|c}{$\widetilde{D_{2}}$} & \multicolumn{1}{|c}{$\mu=\eta/d$} 
\\ \hline
\multicolumn{1}{l}{2d QHE}        &\multicolumn{1}{|c}{2.35} &
\multicolumn{1}{|l}{$0.81\pm.02$} &\multicolumn{1}{|l}{$2.29\pm.02$} &
\multicolumn{1}{|l}{$1.62\pm.02$} &\multicolumn{1}{|l}{$0.80\pm.05$} & 
\multicolumn{1}{|l}{$0.26\pm.05$} \\
\multicolumn{1}{l}{2d Sympl.}     &\multicolumn{1}{|c}{2.75} &
\multicolumn{1}{|l}{$0.83\pm.03$} &\multicolumn{1}{|l}{$2.19\pm.03$} &
\multicolumn{1}{|l}{$1.66\pm.05$} &\multicolumn{1}{|l}{$0.83\pm.03$} & 
\multicolumn{1}{|l}{$0.175\pm.03$} \\
\multicolumn{1}{l}{3d Ortho.}     &\multicolumn{1}{|c}{1.45} &
\multicolumn{1}{|l}{$0.6\pm.05$}  &\multicolumn{1}{|l}{$\sim 4.0$} &
\multicolumn{1}{|l}{$1.7\pm.2$}   &\multicolumn{1}{|l}{$0.55\pm.05$} & 
\multicolumn{1}{|l}{$0.5\pm.1$}   
\end{tabular}
\end{center}
\caption[]{The calculated exponents related to the critical states 
at the metal-insulator transitions in two and three dimensional systems.
The results for the critical exponent of the localisation 
length $\nu$ are taken from \cite{Huc92} (QHE), \cite{Fas92} (2d symplectic), 
and  \cite{HS93,MacK94} (3d orthogonal). The remaining values for 2d systems 
not calculated in the present work were taken from \cite{HS94} (QHE) and 
\cite{Sch95} (2d symplectic).}
\end{table}

\newpage

\newpage

\end{document}